\documentclass[a4paper,fleqn]{cas-dc}
\usepackage[numbers,sort&compress]{natbib}
\usepackage{CJKutf8}

\def\tsc#1{\csdef{#1}{\textsc{\lowercase{#1}}\xspace}}
\tsc{WGM}
\tsc{QE}
\tsc{EP}
\tsc{PMS}
\tsc{BEC}
\tsc{DE}

\usepackage{caption}
\begin{document}
\captionsetup[figure]{labelfont={bf},labelformat={default},labelsep=period,name={Fig.}}

\let\WriteBookmarks\relax
\def\floatpagepagefraction{1}
\def\textpagefraction{.001}
\let\printorcid\relax
% Short title
\shorttitle{}

% Short author
\shortauthors{Xiangqian Zhu et~al.}

% Main title of the paper
\title [mode = title]{Self-supervised inter-intra period-aware ECG representation learning for detecting atrial fibrillation}                      
\author[1,2]{Xiangqian Zhu}

\author[2]{Mengnan Shi}

\author[2]{Xuexin Yu}

\author[2]{Chang Liu}

\author[3]{Xiaocong Lian}

\author[4,5]{Jintao Fei}

\author[4,5]{Jiangying Luo}

\author[1]{Xin Jin}

\author[4,5]{Ping Zhang}

\author[2,3]{Xiangyang Ji}
\cormark[1]

% Address/affiliation
\affiliation[1]{organization={Shenzhen International Graduate School},
    addressline={Tsinghua University}, 
    city={Shenzhen},
    postcode={518000}, 
    country={China}}

% Address/affiliation
\affiliation[2]{organization={Department of Automation},
    addressline={Tsinghua University},
    city={Beijing},
    postcode={100084}, 
    country={China}}
    
\affiliation[3]{organization={Beijing National Research Center for Information Science and Technology (BNRist)},
    addressline={Tsinghua University}, 
    city={Beijing},
    postcode={100084}, 
    country={China}}
    
\affiliation[4]{organization={Cardiovascular Medicine Department},
    addressline={Beijing Tsinghua Changgung Hospital}, 
    city={Beijing},
    postcode={102218}, 
    country={China}}

\affiliation[5]{organization={School of Medicine},
    addressline={Tsinghua University}, 
    city={Beijing},
    postcode={100084}, 
    country={China}}
% Corresponding author text
\nonumnote{*\ Corresponding author.}
\nonumnote{\ \ \textit{E-mail address}: xyji@tsinghua.edu.cn (X. Ji).}

\begin{abstract}
Atrial fibrillation is a commonly encountered clinical arrhythmia associated with stroke and increased mortality. 
Since professional medical knowledge is required for annotation, exploiting a large corpus of ECGs to develop accurate supervised learning-based atrial fibrillation algorithms remains challenging. Self-supervised learning (SSL) is a promising recipe for generalized ECG representation learning, eliminating the dependence on expensive labeling. However, without well-designed incorporations of knowledge related to atrial fibrillation, existing SSL approaches typically suffer from unsatisfactory capture of robust ECG representations. In this paper, we propose an inter-intra period-aware ECG representation learning approach. Considering ECGs of atrial fibrillation patients exhibit the irregularity in RR intervals and the absence of P-waves, we develop specific pre-training tasks for interperiod and intraperiod representations, aiming to learn the single-period stable morphology representation while retaining crucial interperiod features. After further fine-tuning, our approach demonstrates remarkable AUC performances on the BTCH dataset, \textit{i.e.}, 0.953/0.996 for paroxysmal/persistent atrial fibrillation detection. On commonly used benchmarks of CinC2017 and CPSC2021, the generalization capability and effectiveness of our methodology are substantiated with competitive results. 
\end{abstract}

% Keywords
% Each keyword is seperated by \sep

\maketitle

\section{Introduction}
\begin{figure}[h]
\centering
\includegraphics[width=0.9\linewidth]{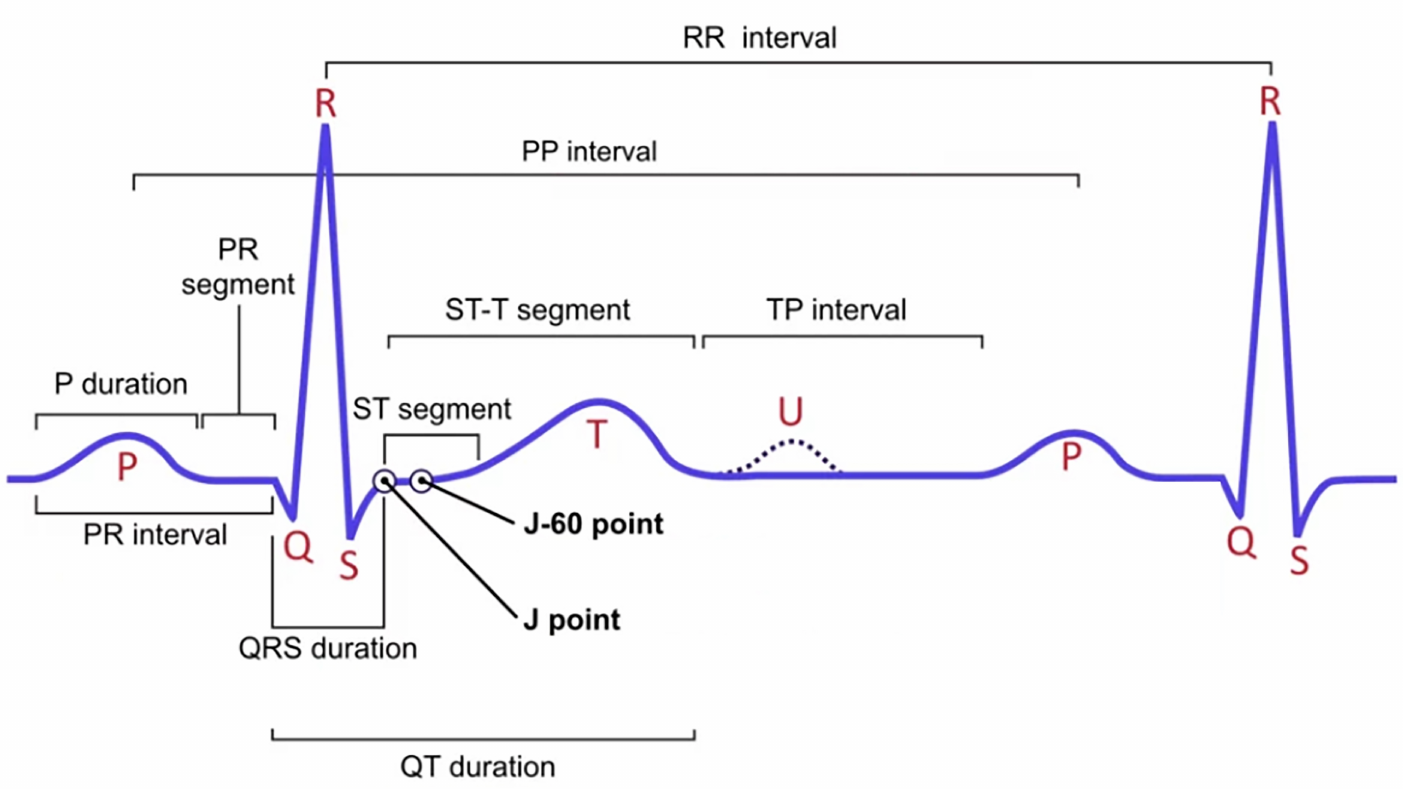} %/fig1pp.png
\caption{The classical ECG curve displays common waveforms and crucial intervals with measurement points.}
\label{fig:ecg curve}
\end{figure}

Atrial fibrillation (AF) is the most prevalent arrhythmia \cite{Kr_l_J_zaga_2022} characterized by irregular heart rate patterns that often begin as paroxysmal episodes and gradually progress to persistent forms.
It is frequently associated with high-risk clinical cardiovascular complications, such as stroke and heart failure, which ultimately lead to increased mortality \cite{Marini_2005,Ott_1997,Wang_2003}. As a common clinical tool for detecting AF, the electrocardiogram (ECG) captures physiological signals that reflect the periodic activities of the human heart \cite{Hagiwara_2018,Lankveld_2014,Tieleman_2014,Serhal_2022}. 
The essential information in the ECG lies within the morphology and intervals (the distance between waves) on the curve \cite{jones2021ecg}, as illustrated in Fig. \ref{fig:ecg curve}. Typically, morphological changes and irregular intervals often indicate corresponding diseases like AF (\textit{e.g.}, the irregularity in RR intervals and absence of P waves). However, the manual interpretation of ECG not only depends on professional knowledge but also introduces an element of subjectivity. This is particularly evident in paroxysmal cases, where subtle features might not be distinctly visible, potentially leading to misinterpretations due to the inherent limitation of human visual perception.
Therefore, developing high-performance algorithms for the automated interpretation and detection of AF in the ECG is of paramount importance and necessity.

Currently, relevant algorithms can be broadly categorized into three types:
\begin{itemize}
\item Handcrafted feature-based algorithms rely on manually extracting features from ECG waves \cite{Couceiro_2008,Logan_2005,Tateno_2001}. Generally, these features relate to P-waves or R-waves obtained manually or through techniques such as symbolization \cite{schafer2015bag}. Subsequently, machine learning classifiers like support vector machines (SVM) or logistic regression (LR) are employed to discriminate and determine the presence of AF. These algorithms mitigate the subjectivity associated with manual ECG interpretation. However, their performance is hindered by the empirical nature of handcrafted feature construction, which struggles to capture rich and complex patterns.

\item Supervised deep learning algorithms automatically extract features \cite{Kulin_2018, Kr_l_J_zaga_2022} from ECG signals. For AF detection tasks, popular deep learning architectures include convolutional neural network (CNN) \cite{Attia_2019,Hannun_2019,Erdenebayar_2019}, long short-term memory (LSTM) networks \cite{Xia_2018}, Transformers \cite{Yan_2019}, and relevant fusion methods \cite{Wang_2020,Limam_2017}. These algorithms offer the advantage of automatically constructing complex features in adaptive high-dimensional spaces but are constrained by the limited availability of large annotated datasets. 

\item Self-supervised learning (SSL) algorithms obtain corresponding representations through pre-training and fine-tuning for downstream tasks. Typically, contrastive learning demonstrates significant potential in ECG representation, involving the construction of a model to assess similarities and differences of pairwise ECG signals, such as mixing up contrastive learning (MCL) \cite{Wickstr_m_2022} and SimCLR \cite{chen2020simple}. In addition, other SSL methods, such as T-S manipulation of temporal-spatial reverse detection \cite{Zhang_2023}, utilize general transformation discrimination pretext tasks. These approaches, leveraging unlabeled data \cite{Morand__2023,Chowdhury_2021,Ohri_2021,Vinh_2023}, present an opportunity to enhance the performance of AF detection. However, existing SSL methods for AF detection often involve simply transferring image or time-series techniques to ECG representations without incorporating medical knowledge, leading to unsatisfactory results, particularly for paroxysmal atrial fibrillation with less prominent features.

\end{itemize}

\begin{figure}[h]
\centering
\includegraphics[width=0.9\linewidth]{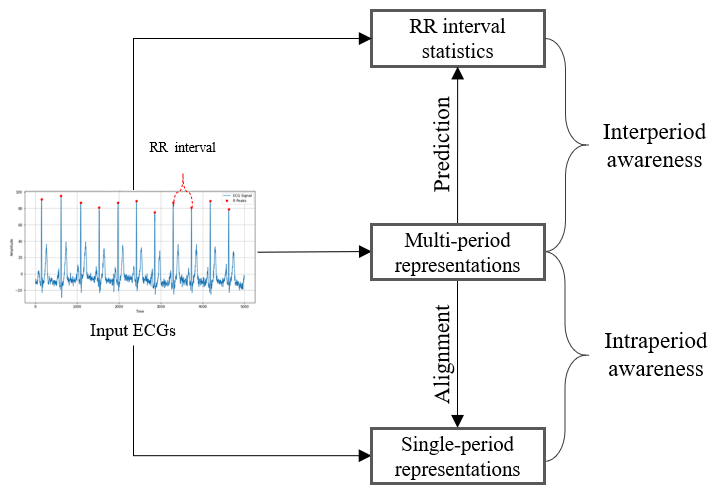} %/fig1pp.png  pipeline.pdf
\caption{Schematic overview of the proposed approach. Inspired by medical knowledge that ECGs of AF patients are related to the irregularity in RR intervals and the absence of P-waves, we propose a self-supervised pre-training method to learn representations with interperiod and intraperiod awareness.}
\label{fig:overview}
\end{figure}

In this study, we propose a novel self-supervised inter-intra period-aware ECG representation learning approach guided by medical domain knowledge, as shown in Fig.~\ref{fig:overview}.
Given that the ECGs of patients with AF are commonly related to the irregularity in RR intervals \cite{duverney2002high,sarkar2008detector} and the absence of P-waves \cite{clavier2002automatic,dotsinsky2007atrial,ehrlich2003prediction}, we design pre-training tasks to learn representations that can capture interperiod variability in the RR intervals and recognize the intraperiod absence of P-waves. Specifically, the interperiod pre-training task encourages the model to perceive periodic representations which facilitates the recognition of the irregularity in RR intervals. In the interperiod task, we first isolate the R-wave positions from the multi-period ECG input and process them independently to calculate the RR intervals, whose statistics, \textit{i.e.}, mean and standard deviation, serve as prediction targets. The task then requires the multi-period encoder to extract representations from the multi-period ECG to accurately predict these targets. Moreover, the intraperiod pre-training task enhances its ability to represent the stable morphology of a single cycle, thereby better capturing the absence of the P-waves. In the intraperiod task, the multi-period ECG is decomposed into multiple single-period segments, then aligned with the primary beats, and the median values are utilized to generate a single-period representative morphology with less noise. Subsequently, we utilize the encoders to extract representations separately and apply contrastive learning to align the multi-period and single-period stable representations from the same ECG record.
Combining the proposed tasks, our model can learn ECG representations from vast amounts of unlabeled data, focusing not only on representative morphologies within individual periods but also on preserving information across multiple periods. Following the acquisition of ECG representations, the model undergoes fine-tuning for AF detection on a relatively small set of labeled target data. 

We focus mainly on paroxysmal atrial fibrillation (AFp) because it is challenging and crucial for early treatment, and conduct preliminary tests on persistent atrial fibrillation (AFf) to demonstrate the effectiveness of our approach. Our approach leverages prior medical knowledge related to AF, including the irregularity of RR intervals and the absence of P waves. After pre-training on a large amount of unlabeled data, our model gains a comprehensive understanding of ECG from both interperiod and intraperiod perspectives. Subsequent fine-tuning on small annotated datasets improves the performance of detection, even in situations of AFp where features are not prominently discernible. This method allows for robust detection by enhancing the distinguishability of relevant features from both interperiod and intraperiod perspectives. Experimental results indicate that the inter-intra period-aware ECG representation learning outperforms previous methods, confirming the significance of period awareness, including both interperiod and intraperiod representations for ECG signals. This holds significant promise for large-scale AF screening and health monitoring, potentially reducing medical professionals' workload and healthcare costs. 
%The contributions of this work can be summarized as follows:
\section{Materials and methods}
\subsection{Dataset description}
Our private dataset is collected from Beijing Tsinghua Changgung Hospital (BTCH) from January 2016 to November 2021. The ECG signals are acquired using GE Healthcare's Marquette ECG device. The device records signals with a duration of 10 seconds and a sampling rate of 500 Hz. These signals consist of 12 leads, where eight of the leads are acquired directly (I, II, and V1-V6), and the remaining four are derived via Einthoven's law (III) and Goldberger’s equations (aVR, aVL, and aVF):
\begin{equation}
\begin{aligned}
& \text { III }= \text{ II }-\text { I } \\
& \text { aVR }=-(\text { I }+\text { II })/2 \\
& \text { aVL }=\text { I }- \text { II}/2 \\
& \text { aVF }=\text { II }-\text { I}/2
\end{aligned}
\end{equation}

The dataset includes a total of 7,676 patients, with a total of 15,695 ECG records. Diagnostic labels are verified on the basis of the results from the Marquette™ 12SL™ ECG Analysis Program and medical history. Ultimately, these labels are annotated as Normal, AFf and AFp by professional physicians. Details are provided in Table \ref {tab:BTCH dataset}.

\begin{table}[width=1\linewidth,cols=4,pos=h]
\caption{Statistical information of the BTCH dataset.}
\label{tab:BTCH dataset}
\begin{tabular*}{\tblwidth}{@{} LLLL@{} }
\toprule
                & Normal       & AFf          & AFp \\
\midrule
Records         & 6735         & 7980         & 980         \\
Subjects        & 4713         & 2617         & 346         \\
Age             & 50.7±17.3    & 73.3±10.8    & 70.3±10.4   \\
Male            & 2723 (40.4\%) & 4763 (59.7\%) & 536 (54.7\%) \\
P-waves absence  & 61 (0.9\%)    & 7804 (97.8\%) & 203 (20.7\%) \\
\bottomrule
\end{tabular*}
\end{table}

To better evaluate the effectiveness of the proposed method, we also utilize two publicly available datasets related to AF: the PhysioNet/Computing in Cardiology Challenge 2017 (CinC2017) \cite{Clifford_2017} and the 4th China Physiological Signal Challenge 2021 (CPSC2021) \cite{wang2021paroxysmal}. CinC2017 dataset is collected by the AliveCor device, with a sampling rate of 300 Hz and a length ranging from 9 s to over 60 s. The complete training set contains 8528 single-lead ECG records categorized into four classes. For our study, we specifically select ECG signals labeled Normal and AF (without distinguishing between AFp and AFf). CPSC2021 dataset is recorded from 12-lead Holter or 3-lead wearable ECG monitoring devices. It provides variable-length ECG records extracted from lead I and lead II of long-term dynamic ECG with a sampling rate of 200 Hz. It consists of 1436 records, extracted from Holter records of 49 AF patients (including 23 paroxysmal AF patients) and 56 non-AF patients. Details are provided in Table \ref {tab:summary}.

\begin{table*}[width=2\linewidth,cols=4,pos=h]
\caption{A summary of the three datasets used.}
\label{tab:summary}
\begin{tabular*}{\tblwidth}{@{} LLLL@{} }
\toprule
               & BTCH          & CinC2017        & CPSC2021\\
\midrule
Normal records  & 6735                              & 5076            & 732                                                      \\
AF records      & 7980 (AFf), 980 (AFp)             & 758             & 475 (AFf), 229 (AFp)                                         \\
Duration        & 10 s                               & 9 s$\sim$60 s      & 8 s$\sim$6 h                                               \\
Sample rate (Hz) & 500                               & 300             & 200                                                      \\
Lead            & I,II,V1$\sim$V6 (III,avR,avL,avF) & /               & I,II                                                     \\
 & & &12-lead Holter \\
Sources         & GE Marquette device               & AliveCor device &  or 3-lead wearable \\
& & &ECG monitoring devices\\
\bottomrule
\end{tabular*}
\end{table*}

\subsection{Signal preprocessing}
\begin{figure*}[h]
\centering
\includegraphics[width=1.0\linewidth]{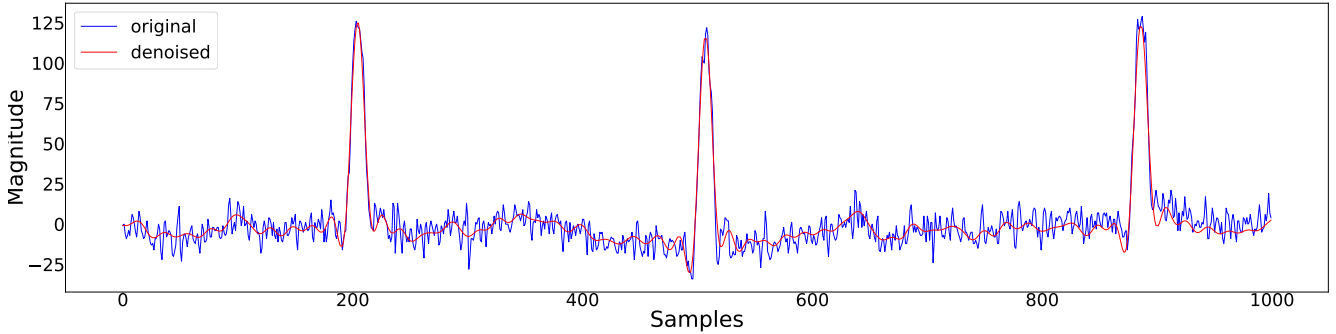}
\caption{Denoising of an ECG signal. The blue line represents the original signal, while the red line represents the signal after denoising. Only 1,000 samples of the ECG signal (representing 2 seconds at a sampling rate of 500 Hz) are displayed here to better distinguish between the original and denoised signals.}
\label{fig:ecg and denoising}
\end{figure*}

In general, the collection of ECG signals is susceptible to several forms of noises such as baseline drift and powerline interference, which make it challenging to realize AF detection precisely and robustly \cite{Uwaechia_2021}. To remove the above noises, we analyze the frequency characteristics of ECG signals by Fast Fourier Transform (FFT) \cite{Aditya_1995}. Based on the fact that ECG signals typically exhibit small frequencies and low amplitudes \cite{Li_2022}, we employ bandpass filtering from 0.5 Hz to 35 Hz and median filtering for further smoothing and denoising. As shown in Fig. \ref{fig:ecg and denoising}, the jagged noise in the original signal is removed, resulting in a relatively clean ECG signal.
Additionally, for the CinC2017 and CPSC2021 datasets, we segment variable-length ECG records into 10-second segments using a sliding window and resample them to 500 Hz. Finally, Z-score normalization is applied to facilitate fast and stable convergence when training models, as shown in Eq. (\ref{eq-normalization}).

\begin{equation}
    \label{eq-normalization}
    X_{\text {normalized }}=\frac{X-\mu}{\sigma}
\end{equation}

\subsection{Pre-training based on period awareness}

\begin{figure*}[h]
\centering
\includegraphics[width=0.8\linewidth]{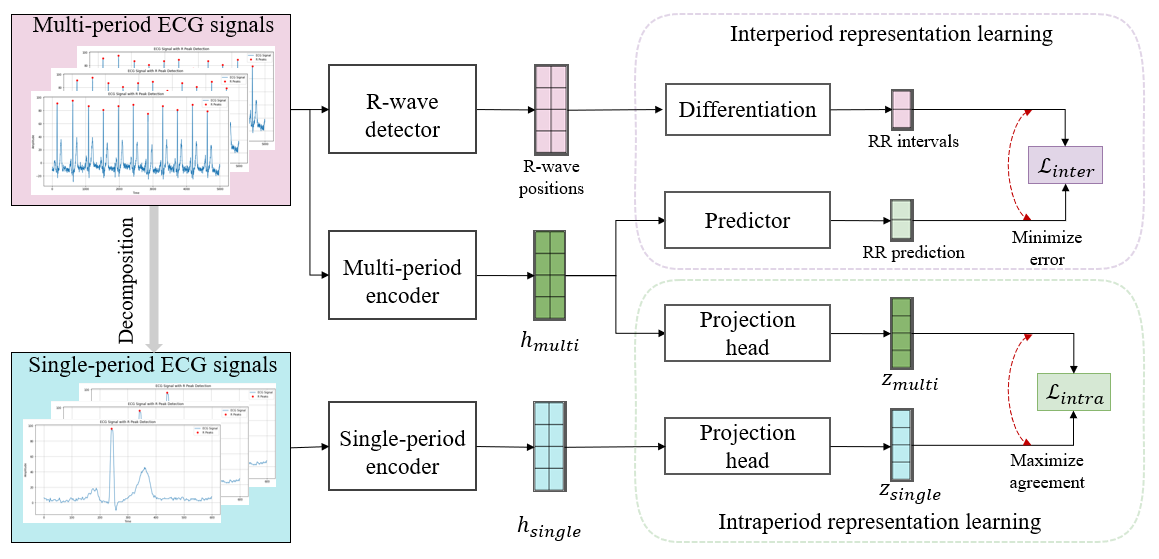} %fig5.png  pretraining.pdf

\caption{An overview of the pre-training architecture for our self-supervised method. The interperiod task primarily involves predicting RR intervals information, while the intraperiod task focuses on aligning single-period morphologies through contrastive learning in this experiment.}
\label{fig:pretrain}
\end{figure*}

To capture effective feature representations for AF detection from large-scale unlabeled ECG data, we propose an inter-intra period-aware ECG pre-training method, as illustrated in Fig. \ref{fig:pretrain}. Based on medical prior knowledge that ECGs of atrial fibrillation patients exhibit the irregularity in RR intervals and the absence of P-waves, we propose our method to capture the stable morphology representation within single-period and valuable representation across periods by exploring interperiod and intraperiod representation, respectively.

\begin{itemize}
    \item Interperiod representation learning: we design an R-wave extractor to determine the absolute positions of R-waves on the time axis. Subsequently, through differential operations, we obtain the RR interval sequence and calculate their mean and standard deviation as prediction targets, which largely reflect the irregularity of RR intervals. For multi-period ECG signals, we develop an encoder to extract features and further predict the above targets. Correspondingly, the mean squared error can measure the loss $\ell_i$ for the $i_{th}$ example, with the optimization objective of minimizing the interperiod loss $\mathcal{L}_{\text {inter }}$ for a minibatch containing $N$ examples, as shown in Eq. (\ref{eq-inter-loss}).
    \begin{equation}
    \label{eq-inter-loss}
    \begin{aligned}
    & \ell_i=\frac{1}{n} \sum_{i=1}^n\left(y_i-\hat{y}_i\right)^2 \\
    & \mathcal{L}_{\text {inter }}=\frac{1}{N} \sum_{i=1}^N \ell_i \\
    \end{aligned}
    \end{equation}
    \item Intraperiod representation learning: Multi-period ECG signals are decomposed into multiple single-period morphologies, then aligned with the primary beats, and the median values are utilized to generate a single-period representative stable morphology, which can dramatically reduce noise. After that, we leverage separate encoders for multi-period and single-period ECG signals, extracting representations denoted as ${h}_{\text{multi}}$  and ${h}_{\text{single}}$, respectively. Employing a projection head, these representations are further transformed into ${z}_{\text{multi}}$ and ${z}_{\text{single}}$ in the latent space. A multi-period representation and its corresponding single-period representation, derived from the same ECG record, constitute a positive example pair, while those originating from different ECG records form a negative example pair. Subsequently, intraperiod representations can be learned by maximizing agreement between via a contrastive loss in the latent space. We randomly sample a minibatch of $N$ examples, comprising $2N$ data points from single-period and multi-period ECGs. Given a positive pair, we treat the other $2(N-1)$ examples within a minibatch as negative examples \cite{chen2017sampling}. Let $\operatorname{sim}(\boldsymbol{u}, \boldsymbol{v})=\boldsymbol{u}^{\top} \boldsymbol{v}/\|\boldsymbol{u}\|\|\boldsymbol{v}\|$ denote the dot product between $\ell_2$ normalized $u$ and $v$ (i.e. cosine similarity). Then the normalized temperature-scaled cross entropy loss $\ell(i, j)$ can be calculated for a positive pair of examples $(i, j)$ \cite{chen2020simple}, with the optimization objective of minimizing the intraperiod loss $\mathcal{L}_{\text {intra }}$ for a minibatch, as shown in Eq. (\ref{eq-intra-loss})
    \begin{equation}
    \label{eq-intra-loss}
    \begin{aligned}
    & \ell(i, j)=-\log \frac{\exp \left(\operatorname{sim}\left(\mathbf{z}_i,   \mathbf{z}_j\right) / \tau\right)}{\sum_{k=1}^{2 N} \mathbb{1}_{[k \neq i]}   \exp \left(\operatorname{sim}\left(\mathbf{z}_i, \mathbf{z}_k\right) /    \tau\right)} \\
    & \mathcal{L}_{\text {intra }}=\frac{1}{2 N} \sum_{k=1}^N[\ell(2 k-1,2  k)+\ell(2 k, 2 k-1)]
    \end{aligned}
    \end{equation}
    where $\mathbb{1}_{[k \neq i]} \in\{0,1\}$ is an indicator function evaluating to 1 iff $k \neq i$ and $\tau$ denotes a temperature parameter to scale the similarity.
    \end{itemize}
    
The total loss of our method combines the above pre-training losses for intraperiod and interperiod representation learning, which is formulated as 
\begin{equation}
\begin{aligned}
\label{eq-all-loss}
 \mathcal{L}_{\text {all }}=\alpha * \mathcal{L}_{\text {intra }}+(1-\alpha) * \mathcal{L}_{\text {inter }}
\end{aligned}
\end{equation}
where $\alpha$ is a hyperparameter used to balance the loss items $\mathcal{L}_{\text {intra }}$ and $\mathcal{L}_{\text {inter }}$.

The R-wave detector adopts the simple \textit{find\_peaks} function from the Scipy library or the sophisticated PanTompkins algorithm \cite{Pan_1985} to obtain positions of R-waves. The supervised learning model is a 34-layer residual network modified from the network architecture\cite{Xie_2017}, which achieves state-of-the-art performances on ECG-related tasks \cite{Hong_2019,Hong_2020}. During the pre-training phase, the first 33 layers serve as the multi-period encoder dropping the last fully-connected layer. 
Simultaneously, a 3-layer convolutional neural network with dilation serves as the encoder for single-period ECG signals. Both the predictor and the projection head are composed of Multilayer Perceptrons (MLPs) to map the representations from encoders to the specified dimension. When pre-training is completed, the task-dependent projection heads are discarded and the encoders are retained for downstream AF detection. 

\subsection{Fine-tuning for atrial fibrillation detection}
\begin{figure*}[h]
\centering
\includegraphics[width=0.8\linewidth]{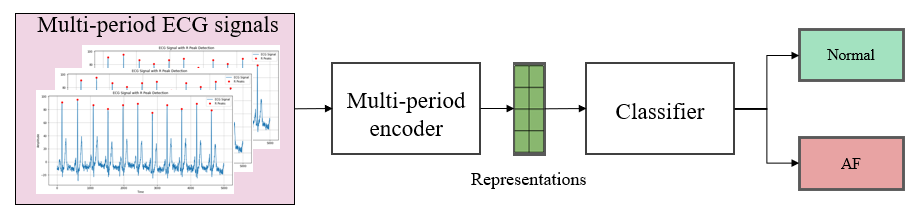} %finetuning.pdf
\caption{The self-supervised fine-tuning process for downstream tasks, specifically atrial fibrillation detection in this experiment.}
\label{fig:fine-tune}
\end{figure*}

During the pre-training, the model's encoders gradually absorb prior knowledge, gaining a comprehensive understanding of ECGs from both interperiod and intraperiod perspectives. Subsequently, the knowledge embedded in the encoder of the pre-trained model is transferred to the downstream model by weight sharing. As illustrated in Fig. \ref{fig:fine-tune}, ECG signals are processed through the encoder and produce refined ECG representations. Subsequently, the ECG representations are fed into a classifier to obtain the probabilities of normal rhythm and atrial fibrillation. Specifically, a fully connected layer outputs two nodes, one for normal rhythm and one for atrial fibrillation, which are then normalized using the Softmax function to yield the respective probabilities. Considering that relevant physiological features, such as sex and age, can provide additional information to enhance atrial fibrillation detection performance, these features can be incorporated with the ECG representations whenever available. Specifically, the physiological features are concatenated with the ECG representations obtained from the encoder and then fed into the fully connected layer for integration and further classification. When assessing the representational capability of the SSL encoder in practical scenarios, two approaches are considered: one involves freezing the weights of the encoder and only updating the weights of the classifier for linear evaluation, while the other entails updating all parameters for full fine-tuning. In our experiments, we exclusively employ the multi-period encoder for both evaluations, highlighting its effectiveness and robustness in learning ECG representation. It is worth mentioning that the single-period encoder is valuable even for datasets consisting solely of single-period data, which enhances the model's scalability and applicability.

\subsection{Visualization and model interpretation}
It is widely recognized that interpretability is crucial for models in medical scenarios, ensuring acceptance and application by healthcare professionals and patients. Unlike traditional approaches that rely on post hoc interpretations to uncover the model's decision rationale, our approach is primarily guided by medical prior knowledge in algorithm design and pre-training tasks. Initially, we leverage clinical expertise to analyze and deduce features associated with AF in ECG signals. Notably, AF may manifest as irregular RR intervals and the absence of P-waves. On the one hand, visualizing the RR interval information within ECG signals confirms its correlation with AF, thereby validating the rationality of its guidance in the task of interperiod representation. On the other hand, statistical analysis of P-waves, which illustrates their association with AF, enabled the decomposition of multi-period signals into single-period representative morphologies. This significantly aids in the extraction of intraperiod features, such as the absence of P-waves. Visualization and statistical analysis have underscored the significance of period awareness and have augmented the interpretability of the proposed pre-training tasks from interperiod and intraperiod perspectives. This ensures that the mechanisms behind our proposed self-supervised learning approach for detecting AF are comprehensible, making it more suitable for clinical acceptance and application.

\section{Experiment details}

\subsection{Implementation details}
% 预训练和微调的参数细节
All models are implemented in Python 3.7 and PyTorch on a server with an NVIDIA GeForce RTX 3090 GPU. During the pre-training phase, the batch size is 16. The Adam optimizer is utilized with an initial learning rate of 0.0003. The scheduler employed is StepLR with a rate decay of 0.9 every 3 epochs, ensuring the stability of the overall optimization process. The entire pre-training process lasts for 30 epochs, which is determined based on the progression of the loss decline in early exploration. During the fine-tuning phase, we use the same initial learning rate as the pre-training phase for the Adam optimizer, and the scheduler with a rate decay of 0.9 every epoch for expedited convergence. A smaller batch size of 4 is adopted, along with a label smoothing rate of 0.1. In addressing the issue of class imbalance in AFp detection, the loss of positive samples is weighted with a factor of 3. This strategy significantly improves the model's performance in scenarios where imbalanced class distribution poses a challenge.

\subsection{Evaluation metrics}
The primary evaluation metric for AF detection is the area under the receiver operating characteristics curve (AUC), which is not affected by sample imbalance and can effectively reflect detection performance. In addition, several auxiliary metrics are employed, including accuracy (ACC), sensitivity (SEN), specificity (SPE), positive predictive value (PPV), and negative predictive value (NPV). Their formulas are as follows:
\begin{equation}
\begin{gathered}
ACC=\frac{TP+TN}{TP+TN+FP+FN} \\
SEN=\frac{TP}{TP+FN} \\
SPE=\frac{TN}{TN+FP} \\
PPV=\frac{TP}{TP+FP} \\
NPV=\frac{TN}{TN+FN}
\end{gathered}
\end{equation}
where, TP represents the number of samples labeled as positive and predicted as positive, TN represents the number of samples labeled as negative and predicted as negative, FP represents the number of samples labeled as negative but predicted as positive, and FN represents the number of samples labeled as positive but predicted as negative.

\subsection{Baseline methods}
% 经典模型简单介绍（BOSS,rocket,SVM,逻辑回归|1DCNN/ResNet1d/XResNet1d/Transformer/）
To comprehensively evaluate our proposed model in AF detection, we compare it against handcrafted feature algorithms and deep learning methods, including supervised and self-supervised approaches. Details of each baseline method are as follows.

(a) Handcrafted feature-based methods: These typically involve manual feature extraction followed by a classifier for detection. BOSS \cite{schafer2015bag} and ROCKET \cite{Dempster_2020} are advanced techniques for extracting manual features. Some traditional machine learning algorithms serve as classifiers in this context, such as SVM and LR. We choose the best result as the final performance of handcrafted feature-based methods from the following four combinations: BOSS + LR, BOSS + SVM, ROCKET + LR and ROCKET + SVM.
 
(b) CNN: CNN is a type of artificial neural network commonly used in ECG classification tasks \cite{Xie_2017}. It consists of multiple layers, including convolutional layers, pooling layers, and the fully-connected layer. The convolutional layers apply convolution operations to extract features such as sharp peaks. The pooling layers downsample the feature maps to reduce computation and control overfitting. Finally, the fully-connected layer combines the features extracted by the convolutional layers to make predictions. 

(c) LSTM: LSTM is a type of recurrent neural network architecture designed to capture long-term dependencies in sequential data. It has a complex structure that includes memory cells, input and forget gates, and output gates. This architecture allows the LSTM network to selectively remember or forget information over long sequences, making it particularly effective for tasks such as physiological signals classification  \cite{shi2015convolutional}. 

(d) Transformer: Transformer is a sequence-to-sequence model that relies on self-attention mechanisms, which allows the model to weigh the importance of different parts of the input sequence when generating an output. The Encoder of Transformer was used for ECG classification, showcasing their potential in capturing long-range dependencies in ECG signals \cite{vaswani2017attention}.

(e) SimCLR \cite{chen2020simple}: SimCLR is a simple framework for contrastive learning of representations. It forms positive sample pairs through two types of data augmentation and maximizes the agreement of these pairs through contrastive loss. Such a pre-training method for learning corresponding representations plays a crucial role in enhancing predictive tasks.

(f) MCL \cite{Wickstr_m_2022}: MCL is an unsupervised contrastive learning framework that is motivated from the perspective of label smoothing. This innovative approach leverages a unique contrastive loss function that seamlessly incorporates a data augmentation strategy. This strategy involves creating new samples by blending two data samples along with a mixing component. The primary objective is to predict the mixing component, which is utilized as soft targets in the loss function. 

(g) T-S \cite{Zhang_2023}: T-S is a simple yet effective self-supervised approach for ECG representation learning. Inspired by the temporal and spatial characteristics of ECG signals, the original signals are flipped horizontally (temporal reverse), vertically (spatial reverse), and both horizontally and vertically (temporal–spatial reverse). Learning is then done by classifying four types of signals including the original one.

To ensure a fair comparison of performance in AF detection, the self-supervised learning models employ the same network architecture. To improve the performance, all deep learning models incorporated basic physiological indicators (sex, age) by concatenating them before the classifier. 
These models are mainly used for the challenging task of AFp detection, due to its lack of distinct features on ECG curves.

\subsection{Ablation study}
% 周期内和周期间和共同
To thoroughly verify the effectiveness of the proposed self-supervised learning strategy, we conduct ablation studies to investigate intraperiod and interperiod representations by comparing the following variants.
\begin{itemize}
\item Interperiod representation detection: $\alpha$ in the total loss $\mathcal{L}_{\text {all }}$ is set to 0, resulting in the model only focusing on interperiod RR interval information related to AF. By exclusively considering the broader temporal context between periods, the model aims to enhance its understanding of patterns that extend beyond individual periods.

\item Intraperiod representation detection: this task entails setting $\alpha$ in the total loss $\mathcal{L}_{\text {all }}$ to 1, emphasizing information within a single period exclusively. Through contrastive learning, the primary objective is to align multi-period ECG signals to stable morphologies within individual periods. 
%Furthermore, by focusing on intraperiod features, the model aims to capture nuanced variations within individual representitive ECG periods.

\item Inter-Intraperiod representation detection: $\alpha$ in the total loss $\mathcal{L}_{\text {all }}$ is set to 0.5, concurrently considering interperiod and intraperiod information. The comprehensive objective is to align multi-period ECG signals to single-period stable morphologies while preserving interperiod differences. This approach seeks to strike a balance, acknowledging the significance of both intraperiod and interperiod representations, aiming for a holistic representation that captures the intricacies of the underlying physiological patterns.
\end{itemize}

\subsection{Scalability experiments}
% 单导联和多导联；数据量
Considering the standard 12-lead ECG, with 8 leads obtained directly, the remaining 4 leads can be calculated using simple formulas. Therefore, our primary experiments focus on the 8-lead ECG signals to reduce computation and speed up training. Subsequently, we extensively conduct experiments involving single-lead and two-lead configurations, in order to validate the potential application of our approach for wearable devices and large-scale health monitoring initiatives. Additionally, we delve into an exploration of the effectiveness of our self-supervised learning approach in comparison to supervised learning, considering diverse data volumes. This comparative analysis aims to provide insights into the scalability and robustness of our methodology across different datasets, shedding light on its potential for real-world applications in various healthcare scenarios.

\section{Results}
\subsection{Results of baseline methods}

Table \ref{tab:comparison result} presents the comparative results of baseline methods and our self-supervised inter-intra period-aware ECG representation learning for AFp detection on the BTCH dataset. Despite differences in feature extraction and classification methods, handcrafted feature algorithms tend to show lower AUC and sensitivity. Although Transformer and LSTM models are adept at processing general sequence data, their effectiveness is limited in the context of multi-lead and multi-period ECG signals, where they marginally outperform handcrafted features. CNN  excels in identifying morphological variations in ECG signals, resulting in superior disease detection capabilities with a competitive AUC score and high sensitivity. The performance of these supervised learning methods is constrained by the labeled data, making it difficult to achieve further improvement.  

Furthermore, we compare it with existing popular self-supervised methods, including SimCLR \cite{chen2020simple}, MCL \cite{Wickstr_m_2022}, and T-S \cite{Zhang_2023} methods. To objectively evaluate the performance of models, the final results of SimCLR are the average outcomes derived from the following two pairs of augmentation methods: (1) adding noise and random permutation; (2) horizontal flipping and vertical flipping. After completing the pre-training task, we load the pre-trained model weights and perform full fine-tuning for AF detection. As shown in Table \ref{tab:comparison result}, self-supervised learning methods generally outperform supervised learning methods because self-supervised learning leverages unlabeled data to mine more information. Besides, the proposed inter-intra period-aware ECG representation method outperforms existing self-supervised learning methods, achieving outstanding results with an AUC of 0.953 and sensitivity of 0.854. The SimCLR and T-S exhibit improvement when utilizing horizontal flipping and vertical flipping transformations. 
MCL employs the mixture of two data samples with hyperparameter-adjusted similarity, which also yields favorable results. This means that after pre-training of these transformations, models are able to capture features beneficial to downstream AF detection.  

\begin{figure}[h]
\centering
\includegraphics[width=1\linewidth]{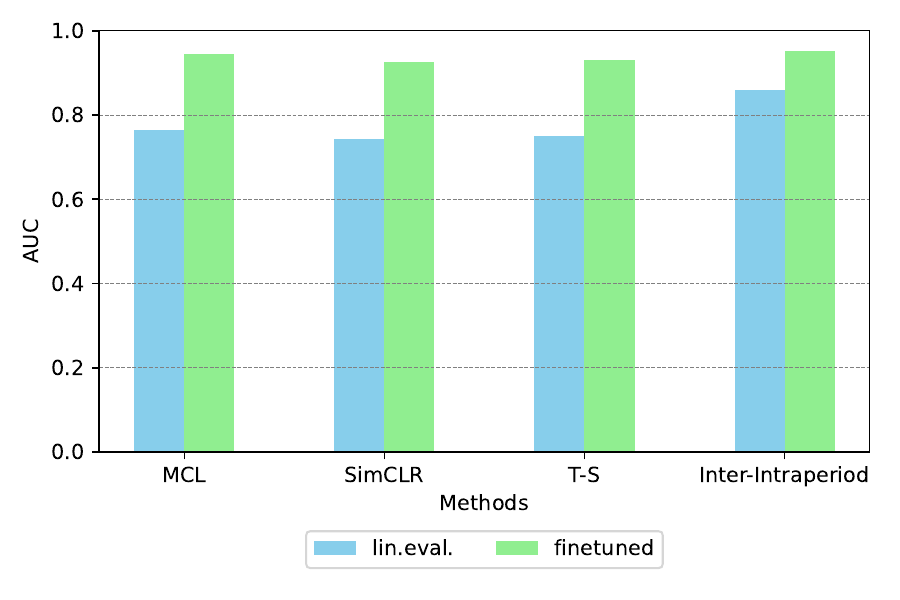}

\caption{Comparison of AUCs between other self-supervised learning and our method. The blue bars represent linear evaluation, while the green bars represent full fine-tuned results.}
\label{fig:AUC line&fine-tune}
\end{figure}

Additionally, we conduct linear evaluations of the capacity AF detection on the models after pre-training. As depicted in Fig. \ref{fig:AUC line&fine-tune}, despite all methods performance not matching that of full fine-tuning, our method outperforms previous self-supervised methods in linear evaluation and exhibits a significant advantage in the pre-training task for ECG signals, which further verifies the effectiveness of our intraperiod and interperiod representations.

To validate the efficacy of our self-supervised learning method, we conduct experiments on publicly available datasets, specifically CPSC2021 and CinC2017, as shown in Table~\ref{tab:public result}. The CPSC2021 dataset distinguishes between paroxysmal atrial fibrillation (AFp) and persistent atrial fibrillation (AFf), while the CinC2017 dataset does not differentiate between types of atrial fibrillation (AF). On the CPSC2021 dataset, our proposed method outperforms existing self-supervised learning methods for both paroxysmal and persistent atrial fibrillation. The detection performance for persistent atrial fibrillation is better than for paroxysmal atrial fibrillation because persistent atrial fibrillation exhibits more distinct features. Even on the CinC2017 dataset, which does not differentiate between types of atrial fibrillation, our method demonstrates excellent detection performance, confirming the effectiveness and generalization capability.

\begin{table*}[width=2\linewidth,cols=8,pos=h]
\caption{The comparison of our model with others in detecting paroxysmal atrial fibrillation on the BTCH dataset.  The bold letters in the table indicate the best result in that column, while underscores represent the second-best result.(Ditto for other tables.)}
\label{tab:comparison result}
\begin{tabular*}{\tblwidth}{@{} LLLLLLLL@{} }
\toprule
  & Methods& {\textbf{AUC}} & ACC & SEN & SPE & PPV & NPV \\
\midrule
\multirow{4}{*}{Supervised learning}&
Handcrafted \cite{schafer2015bag}       & 0.805                           & 0.875                  & 0.030                    & \textbf{0.999}     & \textbf{0.857}          & 0.875                   \\
&LSTM \cite{shi2015convolutional}            & 0.766                            & 0.771                   & 0.282                   & 0.846              & 0.220                   & 0.884                   \\
&Transformer \cite{vaswani2017attention}     & 0.916                      & 0.878                 & 0.768                   & 0.893              & 0.503                   & 0.965                   \\
&CNN \cite{Xie_2017}      & 0.912                            & 0.804                   & \textbf{0.883}          & 0.792              & 0.396                   & \textbf{0.978}          \\

\midrule
\multirow{3}{*}{Self-supervised learning}&
MCL \cite{Wickstr_m_2022}               & \underline{0.946}    & 0.907                  & 0.821    & 0.919                   & 0.586                   & 0.973             \\
&SimCLR \cite{chen2020simple}          & 0.927          & 0.898                   & 0.753          &  0.921          & 0.600          & 0.960                   \\
&T-S \cite{Zhang_2023}              & 0.931          & 0.899                   & 0.767          & 0.919             & 0.594            & 0.962   \\
\midrule
\multirow{3}{*}{Ablation study}
&Interperiod       & 0.939          & \textbf{0.921}         & 0.796          & \underline{0.940}            & \underline{0.672}     & 0.968                   \\
&Intraperiod       & 0.945    & 0.889                   & 0.767          & 0.907                   & 0.560                   & 0.962                   \\
&Inter-Intraperiod (ours) & \textbf{0.953} & \underline{0.912}                   & \underline{0.854} &  0.921                   & 0.624                   & \underline{0.976 } \\

\bottomrule
\end{tabular*}
\end{table*}

\subsection{Results of ablation study}
\begin{figure}[h]
\centering
\includegraphics[width=1\linewidth]{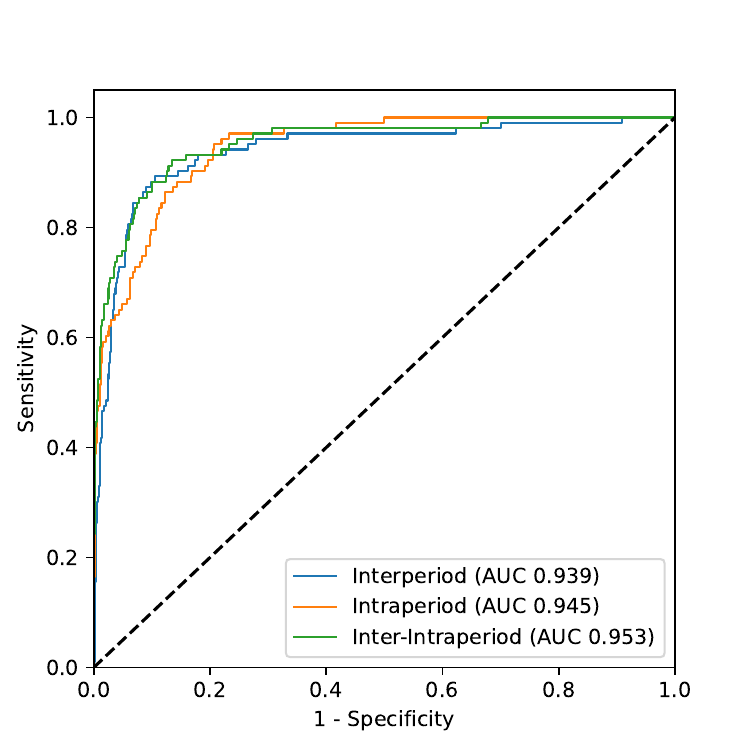}
\caption{Receiver Operating Characteristic (ROC) curves for ablation study.}
\label{fig:ROC}
\end{figure}

Table \ref{tab:comparison result} and Fig. \ref{fig:ROC}  show the results of the ablation study for AFp detection on the BTCH dataset. The results illustrate that intraperiod representation exhibits relatively superior interperiod representation in AUC. This disparity may arise from the relatively simplistic nature of the interperiod task, which concentrates solely on RR interval information. In contrast, the intraperiod representation method captures a more extensive set of morphological information, providing a richer context throughout the entire stable period in ECG signals. Meanwhile, the performance is best when intraperiod and interperiod representations are combined, which indicates both representations capture complementary features. 

\begin{table*}[width=2\linewidth,cols=7,pos=h]
\caption{The AUC results compared with current self-supervised learning methods on two publicly available datasets. CPSC2021 further distinguishes between paroxysmal atrial fibrillation (AFp) and persistent atrial fibrillation (AFf), while CinC2017 does not make further distinctions for atrial fibrillation (AF).}
\label{tab:public result}
\begin{tabular*}{\tblwidth}{@{} LLLL@{} }
\toprule
Methods          &CPSC2021(AFp)  &CPSC2021(AFf ) &CinC2017(AF)\\
\midrule
MCL\cite{Wickstr_m_2022}	                   &\underline{0.931}   &0.983    &0.976				\\ 
SimCLR \cite{chen2020simple}	               &0.912   &0.978    &0.964				\\
T-S	\cite{Zhang_2023}                   &0.925   &\underline{0.987}    &\underline{0.981}				\\
Inter-Intraperiod(ours)&\textbf{0.946}   &\textbf{0.999}    &\textbf{0.991}				\\
\bottomrule
\end{tabular*}
\end{table*}

\subsection{Results of scalability experiments}

We conduct experiments using single-lead (lead I) and two-lead (lead I, II) configurations, aiming to explore the potential application of the proposed method in wearable devices and large-scale health monitoring with reduced data acquisition costs.
As shown in Table \ref{tab:partitional leads result}, the application of our self-supervised learning method on single-lead ECG demonstrates an improvement compared to pure supervised learning. Furthermore, it reveals that the improvement in performance from using our self-supervised learning exceeds that of adding an extra lead (AUC, 0.916 vs. 0.899), affirming the efficacy and effectiveness of using fewer leads with self-supervised learning for AF detection.

\begin{table*}[width=2\linewidth,cols=8,pos=h]
\caption{Comparison results between supervised model and our Inter-Intraperiod (fine-tuned model) on single-lead and two-lead ECGs.}
\label{tab:partitional leads result}
\begin{tabular*}{\tblwidth}{@{} LLLLLLLL@{} }
\toprule

 & & {\textbf{AUC}} & ACC & SEN & SPE & PPV & NPV \\
\midrule
\multirow{2}{*}{Supervised model}           & {I}   & 0.894       & {0.810} & 0.845          & 0.804          & {0.399} & 0.971 \\
                                               & I+II      & 0.899       & 0.883      & {0.699} & {0.912} & 0.550        & 0.952 \\
\multirow{2}{*}{Inter-Intraperiod} & I         & {0.916} & 0.858      & {0.835} & {0.861}    & 0.480        & 0.971 \\
                                               & I+II      & 0.918       & 0.890       & 0.767          & 0.909          & 0.564       & 0.962\\
\bottomrule
\end{tabular*}
\end{table*}

Additionally, we compare the performance of our self-supervised learning method and supervised learning method trained from scratch under different data volumes. As shown in Table \ref{tab:diff num result}, our self-supervised method generally outperforms the supervised learning method trained from scratch across varying data volumes. Notably, the performance improvement of our self-supervised learning method is particularly pronounced when dealing with limited labeled data.

\begin{table*}[width=2\linewidth,cols=7,pos=h]
\caption{AUCs for the model trained from scratch and for our self-supervised learning model using Inter-Intraperiod representations with different numbers of training data.}
\label{tab:diff num result}
\begin{tabular*}{\tblwidth}{@{} LLLLLLL@{} }
\toprule
 \# of training records             & 100 & 200   & 500        & 1000           & 2000           & 5000        \\
\midrule
From scratch (AUC) & 0.534  & 0.564 & 0.620 & 0.720           & 0.844          & 0.925 \\
Inter-Intraperiod (AUC)  & 0.712        & 0.727 & 0.772      & 0.798 & 0.858 & 0.946  \\
\bottomrule
\end{tabular*}
\end{table*}

\subsection{Visualization and model interpretation}
\begin{figure}[h]
\centering
\includegraphics[width=0.8\linewidth]{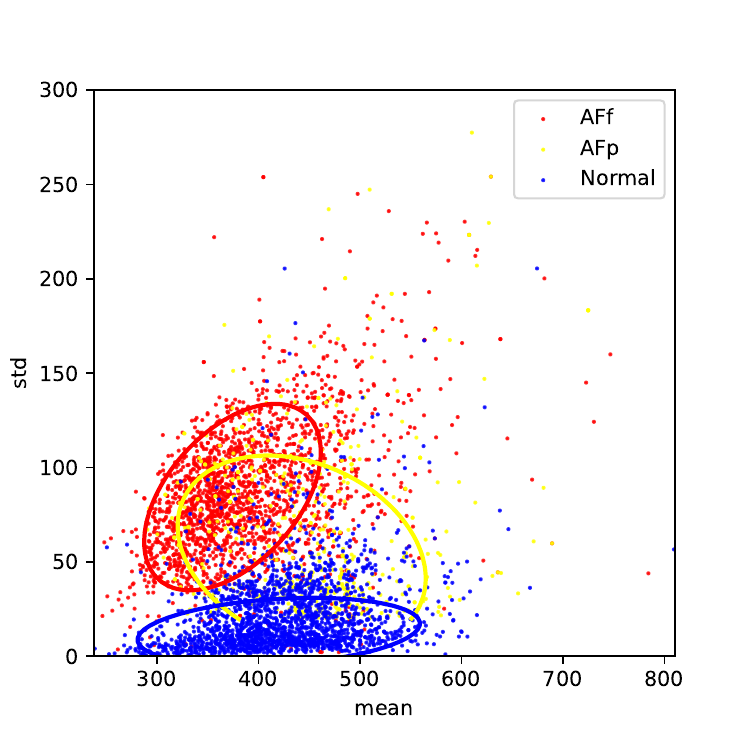}  %fig10.png
\caption{Distribution of RR intervals. The horizontal axis represents the mean, while the vertical axis represents the standard deviation. The elliptical contours enclose 80\% of the points for the corresponding category.}
\label{fig:RR}
\end{figure}

%【图9-RR分布】【图10-tSNE】
In order to enhance the interpretability of our method used for AF detection, we conduct a series of corresponding visualizations and analyses about interperiod and intraperiod representations. In the context of interperiod representation, Fig. \ref{fig:RR} illustrates the distribution of mean and standard deviation of RR intervals based on the BTCH dataset. We could observe clear differences between normal and AFf, while the distinctions related to AFp are relatively subtle, contributing to the greater challenge and value in AFp detection. For intraperiod representation, we conduct a statistical analysis of the absence of the P-waves feature. The findings indicate that AFf exhibits the absence of P-waves in 97.8\% of cases, whereas AFp shows the absence of P-waves in only 20.7\% of cases. In normal individuals, the absence of P-waves in a very rare few cases, approximately 0.9\% of instances. These confirm the rationality of prior knowledge used to guide pre-training tasks from the perspectives of interperiod and intraperiod representations.

\section{Discussion}

Our self-supervised learning method conducts ECG representations from interperiod and intraperiod perspectives. Self-supervised pre-training is employed to obtain ECG representations from massive unlabeled data, overcoming the time-consuming and labor-intensive challenges of data annotation. The design of the pre-training task is guided by medical expertise, recognizing that AF patients often exhibit the irregularity RR intervals and the absence of P-waves in their ECG signals, corresponding to interperiod and intraperiod features, respectively. Further, fine-tuning is performed with a small amount of annotated data to enhance the performance of downstream AF detection.

Subsequently, the models are fine-tuned on a relatively small labeled dataset, primarily focusing on AFp detection. As shown in Fig. \ref{fig:ROC} and Table \ref{tab:comparison result}, both interperiod and intraperiod pre-training tasks make substantial contributions. The amalgamated representation of interperiod and intraperiod features notably enhances AF detection, surpassing the performance of existing self-supervised learning methods.  Additionally, experiments on ECG of partial leads confirm the effectiveness of using fewer leads with self-supervised learning for AF detection. These offer potential support for wearable devices and large-scale health screenings.  Table \ref{tab:diff num result} shows that compared to training from scratch, our self-supervised learning method enhances performance, particularly in scenarios with limited data. 

In future work, we plan to explore the representation of other cardiovascular diseases in ECG data based on medical prior knowledge, guiding pre-training tasks to improve the detection capabilities for specific conditions. Additionally, we aim to collect ECG data from wearable devices and evaluate their effectiveness in detecting AF, particularly in the context of early screening and large-scale health monitoring, with the expectation of simplifying medical procedures.

\section{Conclusion}

In this study, we propose a self-supervised inter-intra period-aware ECG representation learning method specifically designed for detecting atrial fibrillation (AF). Our approach consists of two key stages: pre-training on a large unlabeled dataset to learn robust ECG representations, and fine-tuning on a relatively small labeled dataset for AF detection.
Guided by medical prior knowledge, we design pre-training tasks that capture periodic features of ECG signals. This enables our model to develop a comprehensive understanding of ECG patterns from both interperiod and intraperiod perspectives. During the fine-tuning stage, these learned representations are further refined to enhance their suitability for AF detection.
Our experimental results demonstrate that the proposed method outperforms existing self-supervised approaches in terms of AF detection performance. The ablation study confirms the importance and effectiveness of incorporating both interperiod and intraperiod representation learning. Additionally, leveraging medical knowledge not only improves model performance but also enhances interpretability, making our model more clinically acceptable.
Furthermore, we assess the scalability of our model on partial-lead ECG signals, confirming its superiority in configurations with a small number of leads, with the expectation of reducing data collection costs. This suggests potential applications in wearable devices and large-scale health monitoring systems, offering prospects for reducing healthcare costs through efficient AF detection.
In summary, our method presents a significant advancement in self-supervised ECG representation learning, effectively combining medical insights with advanced machine learning techniques to deliver a robust and viable solution for AF detection.

%% Loading bibliography style file
\bibliographystyle{model1-num-names}
% \bibliographystyle{cas-model2-names}
% \bibliographystyle{plain}

% Loading bibliography database
\bibliography{cas-refs}

\vskip3pt

\end{document}